## Research Article

# Coimagining the Future of Voice Assistants with Cultural Sensitivity

Katie Seaborn, Yuto Sawa, and Mizuki Watanabe

*Department of Industrial Engineering and Economics, Tokyo Institute of Technology, Tokyo 152-8550, Japan*

Correspondence should be addressed to Katie Seaborn; seaborn.k.aa@m.titech.ac.jp

Voice assistants (VAs) are becoming a feature of our everyday life. Yet, the user experience (UX) is often limited, leading to underuse, disengagement, and abandonment. Co-designing interactions for VAs with potential end-users can be useful. Crowdsourcing this process online and anonymously may add value. However, most work has been done in the English-speaking West on dialogue data sets. We must be sensitive to cultural differences in language, social interactions, and attitudes towards technology. Our aims were to explore the value of co-designing VAs in the non-Western context of Japan and demonstrate the necessity of cultural sensitivity. We conducted an online elicitation study ($N = 135$) where Americans ($n = 64$) and Japanese people ($n = 71$) imagined dialogues ($N = 282$) and activities ($N = 73$) with future VAs. We discuss the implications for coimagining interactions with future VAs, offer design guidelines for the Japanese and English-speaking US contexts, and suggest opportunities for cultural plurality in VA design and scholarship.

## 1. Introduction

Voice-based interaction is entering our daily lives in the form of intelligent agents, conversational interfaces, smart speakers, and other interactive forms of artificial intelligence (AI) [1–4]. These voice assistants (VAs) and voice user interfaces (VUIs) offer voice input and output (I/O), a more natural form of interaction that is hands-free and multimodal [1, 4]. Reports on the global VA market reveal a substantial industrial investment and consumer engagement. For instance, sales data on smart speaker adoption from the US in 2020 indicate a 2.9-fold increase from the year before (255 million versus 87.7 million) (https://voicebot.ai/2021/04/14/u-s-smart-speaker-growth-flat-lined-in-2020/). In Japan, 5.8 million smart speakers were bought in 2020, with market trends forecasting 15 million by 2025 (https://www.statista.com/statistics/1024353/japan-smart-speaker-household-penetration). The mainstreaming of VAs in daily life raises the question of what kinds of services and interactions VAs *should* offer and for *whom*.

Notably, language and culture must be considered in the design of voice-based technologies like VAs. VAs require natural language processing (NLP) and data sets in the language of users. Cultural features of language use, the "social" aspects of the experience, and attitudes towards technology can all impact VA acceptance [5, 6]. Yet, most NLP data sets used to train VAs are of English origin and/or based in Western cultural contexts, primarily the US [7]. Critical scholarship [8–11] has warned against this "WEIRD" sampling bias, referring to the overreliance on sampling from Western, Education, Industrial, Rich, and Democratic nations [12] and treating these samples as representative of all people, which a wealth of cross-cultural research has refuted [9, 12, 13]. Japan is a case study in reliance on Western VA imports, like Amazon Alexa, and Apple Siri.

Within and beyond Japan, there is a dire need to improve the user experience (UX) of VAs. Common issues include poor speech recognition and conversational ability [1, 14, 15], limited activities and poor fit with different user groups [4, 16–18], and a misguided focus on commercial functionality [19]. A critical barrier specific to VAs is how to start and end the interaction using voice alone: the *dialogic* design. Users may not know the "wake words" or accidentally "wake up" a VA that "overhears" its name.

Breakdowns centre around VAs being unable to "understand" user dialogue and requests, which is the most common reason for disuse [1, 3, 20, 21]. Moreover, end-users desire and can imagine a broader range of interactions with VAs: the *activity* design. VAs commonly offer a limited set of features that tend to be transactional [1, 15, 20, 22]. Yet, Woodward et al. [18], for example, found that kids wanted to converse at length with VAs socially, without machinelike prompts. Other factors may play a role, notably general attitudes towards technology [15, 22], which could be based in cultural attitudes or market differences across nations. In short, we must consider *how* to design VAs *for the user*, present and future, wherever they are in the world.

Previous work has shown the value in co-design [1, 15–18, 23, 24]. Co-design is a design approach where "the creativity of designers and people not trained in design [work] together" ([25]:6) to imagine, prototype, test, and sometimes develop technologies. Co-design can offer insights that spark new ideas grounded in user needs and mental models [25]. In VA design, *crowdsourcing* is an emerging solution [15–17, 23, 24, 26] for translating and "coimagining" new NLP data sets, for example, Amazon's MASSIVE project [27]. Still, these efforts have been limited to creation for or translation from a single lingo-cultural context, usually "WEIRD" American English. As such, a culturally sensitive approach to co-designing future interactions with VAs is a logical and necessary next step.

To this end, we conducted an online crowdsourced cross-cultural elicitation study on ideal forms of engagement with VAs. We asked Japanese speakers in Japan and English-speaking Americans to write dialogues and ideas for activities. The US represents a typical WEIRD nation [9, 12] and is also where most NLP data sets for VAs have been created. Japan, being Eastern, is only one point of departure away from "W"EIRD. It is also a context for which US- and English-based data sets have been translated for VAs. All else being equal, this allowed us to isolate broad linguistic and cultural differences. Our overarching research question (RQ) was: *What kinds of interactions do American and Japanese people imagine with ideal VAs?* We aimed to reveal cultural differences in these visions, especially based on linguistic expressions, dialogic structures, and social norms, but also recognizing potential cultural differences in attitudes towards technology [1, 15]. Our main contributions are as follows:

(i) Empirical evidences that envisioned interactions with VAs, specifically dialogues and activities, vary across language, culture, and technology attitudes

(ii) An emerging set of guidelines and implications that are cross-cultural and sensitive to the Japanese context

(iii) Our English and Japanese data sets (https://osf.io/bm8r2), translated into both languages

This work advances the state of voice UX in a culturally sensitive way.

## 2. Related Work

*2.1. Co-Designing VAs: Benefits, Drawbacks, and Opportunities.* Despite uptake, a core question remains: what kinds of experiences do people want with VAs? One way to answer this question is to *involve* people in the design process. Indeed, co-design is a staple of human-computer interaction (HCI) research and design [25, 28, 29] that is starting to be explored for VAs [17, 18, 24, 30–32] and other socially interactive AI [16, 33]. Co-design methods have also been used for conversational exchanges, linguistic expressions, and dialogic interaction [34]. The value lies in gathering insights on and brainstorming possibilities with potential and actual end-users. This can help experts understand people's *current* mental models [16, 18, 32, 35], i.e., "VAs can search things online for me," as well as what they can *imagine*, i.e., "I could take my VA on a walk and talk about the history of the park." Garg and Sengupta [16], for instance, asked kids to imagine VA personas. Kids imagined VAs that adapted to different contexts and were emotionally intelligent, able to understand and respond to the child's emotional state. VAs of the present may be wanting, but people of all ages can imagine other possibilities.

In NLP [23, 36, 37], co-design centres around crowdsourcing: gathering large numbers of people who complete microtasks towards a common goal. Crowdsourcing has been used to gather utterances and dialogues within particular contexts of use [15, 36, 38], translate and localize existing data sets [39, 40], and elicit reactions to voice stimuli, especially social and paralinguistic characteristics [41, 42]. A notable example is the translation and localization of the Amazon MASSIVE data set into 51 languages with Amazon Mechanical Turk [27]. This signals a shift in how co-design is being approached: from the classical model of small-scale focus groups and jams to a larger-scale, online, crowd-driven model that has global reach.

Elicitation methods may be able to capture contextualized responses through scenario-based prompts. Reicherts et al. [43], for instance, provided eight scenarios to elicit ideas on potential applications of smart speakers. Similarly, we used scenarios to help respondents think outside of the box. Elicited conversations can also reveal expected or desired features in VA behaviour, such as longer conversations, emotional intelligence, and agent personality. Elicitation methods are also a feasible and scalable co-design approach, able to be conducted online and crowdsourced [15, 36, 38]. Völkel et al. [15], for instance, conducted a crowdsourced dialogue elicitation study to assess how people envision conversations with an ideal voice assistant. They identified several patterns in how this engagement was characterized. We use this work as a basis for our cross-cultural research design and analysis.

Coimagining future VAs with nonexperts can generate useful data on user mental models. However, there are limits to consider and critically examine (our RQ3). One is the tendency for people to rely on current models and stereotypes. Recent work on VAs like Alexa and Siri have found compelling evidence on the user *and* expert sides with respect to stereotyped and even hostile framings of VAs [44–46]. This is not unprecedented; decades of work on computer agents, notably the computers are social actors (CASA) model [47], have shown that we tend to react to humanlike cues—even small ones—in line with our models of the human world. We are generally unaware of it, and our brains often lean on the simplest models available, i.e., stereotypes. This seems to apply to VAs, as well [4]. We therefore took a critical and reflexive frame [48] when evaluating the responses. We analyzed patterns related to people, notably gender, as well as machines, notably scorn and abusive conduct [49]. However, we avoided removing "inappropriate" material, such as swear words [27, 36]. These forms of exchanges need to be trained into VAs so that VAs can recognize and respond appropriately [37]. They may also be natural and appropriate from the end-user's perspective [50]. For example, queer language may be deemed "toxic" in some contexts but represents a takeback of power through language in queer contexts [51]. As such, we did not remove "inappropriate" content, only gibberish, and analyze this content with the rest of the data.

Nonexperts may also have varying levels of experience with VAs and technology in general. In line with Clark et al. [1], we also realized that attitudes towards technology could be a mediating factor. As such, we explored the role of technology attitudes in our analyses.

*2.2. Approaching the Co-Design of VAs Cross-Culturally.* Critical voices have called out bias in who is "the user" [10, 52, 53]. Notably, a range of work [9–12] has identified widescale WEIRD sampling biases. Findings from WEIRD populations are taken as generalizations about humanity and mainstreamed. Yet, research with non-WEIRD populations has revealed differences and culturally sensitive insights [9]. The link between Western nations and English sampling biases [54] suggests that work on VAs may be even WEIRDer. We thus targeted English-speaking participants in the US and Japanese-speaking participants in Japan. Both nations are strong industrially, especially in technology. The US is a typical oversampled Western nation [9, 12], while Japan differs by only one letter on the WEIRD spectrum, i.e., language and culture. Notably, translations of English and Japanese NLP data sets are common, such as the crowdsourcing initiatives of Tatoeba (https://tatoeba.org/en/downloads) and MASSIVE [27]. Yet, biases have been found within these data sets: "missteps" resulting from the crowdsourced translation process [55]. This suggests that translation may be insufficient. Instead, we may need to co-design new data sets with people to capture the particularities of language and culture. However, this has not been explored. We do so in this work, providing evidence by comparing co-designed Japanese and US data sets.

In fact, almost no cross-cultural work on VA design exists. Ma et al. [6] conducted an online survey with people in Germany, Egypt, and China about emotionally aware VAs. They identified three types of cross-cultural user orientations: enthusiasts, pragmatists, and skeptics. Huang and Zhang [5] considered how Taiwanese and UK customers differed on preferences for VA interactions. They found opposite patterns for media richness. In transactional situations, British participants preferred low media richness, i.e., simple, plain, and perhaps blunt expressions. In nontransactional situations, however, they preferred high media richness, i.e., language variety and extra information. The opposite was true for those in Taiwan. In Japan, almost no work to date has explored dialogic interaction with VAs. Ouchi et al. [56] found that plain language was preferred over polite language in the context of making travel plans. In sum, cross-cultural work can reveal cultural similarities and differences, as well as highlight the special features of a given culture or language context.

We may draw inspiration from work in the NLP space, such as MASSIVE, that is now attempting to address language and cultural gaps. However, these initiatives are limited by their starting point: English and the American context [7, 36]. Translation and localization are often not one-to-one, word-for-word, or even phrase-to-phrase processes. In Hoft's iceberg model of culture [57], what is known on the surface, i.e., expressed in dialogue, may not represent the whole picture. Unspoken and unconscious rules and social norms exist [57, 58]. Biases present in the data sets may be retained on translation. For instance, implicit language biases related to gender have been found in the English MASSIVE data set [59]. At present, it is unknown if these biases have been retained in the 50 translations. Finally, the data sets in NLP are geared around dialogue. To the best of our knowledge, there have been no crowdsourcing efforts on imagining activities with VAs. Yet, smaller, face-to-face co-design studies have shown that people can imagine new activities for VAs [16–18, 30]. Crowdsourcing online could scale up this co-design task and generate more ideas. We thus decided on a culturally sensitive elicitation approach using crowdsourcing, asking people in their own language and cultural milieu to generate dialogues and activities for us.

## 3. Methods

We conducted an online elicitation study using the dialogue elicitation method of Völkel et al. [15]. Participants were presented with a range of scenarios to prompt dialogue creation and brainstorm activities with an imagined VA in a questionnaire format. For each scenario, participants wrote an original dialogue, line by line, between an imagined VA and themselves ("you"). We asked them to imagine an "ideal" VA, defining "ideal" however they desired. We extended the approach of Völkel et al. [15] by also asking participants to imagine activities with VAs, without considering the technical standards of today. As in Völkel et al. [15], this protocol combined established elicitation methods in HCI, e.g., ideating new forms of gestures [60] with a narrative format, where participants write a story of the interaction [61]. This method is culture free: stories and brainstorming are human-wide activities.

Before the main study, we conducted a pilot test in-lab ($n = 6$) of the English and Japanese questionnaires. We registered our protocol on OSF (https://osf.io/heurc) before data collection on October 22, 2021. Data was collected in two phases: on November 25, 2021, and November 10, 2022. This was to account for a randomization error in our research platform when assigning participants, which resulted in uneven numbers of Japanese and American respondents.

*3.1. Participants.* Participants were recruited online using SurveyMonkey Audience, which provides a level of quality roughly equivalent to within 10% of traditional market surveys [62]. SurveyMonkey also guarantees the demographics and location of participants, e.g., regularly profiling eligible participants in Japan, who receive compensation in yen, etc. (https://www.surveymonkey.com/market-research/data-quality). A total of 135 valid submissions were received (Table 1). Not all data could be used due to missing or unclear (e.g., gibberish) responses. Participants were compensated with roughly USD $8 or JPN ￥800 for one hour. This study was approved by the university ethics board and conducted in accordance with the Declaration of Helsinki (1964).

*3.2. Scenarios.* We created three scenarios of escalating complexity to prompt dialogue creation. The first was an everyday VA scenario, according to research on transactional VA use, e.g., [15, 21, 63]: "You want to talk about the weather forecast on the weekend" (S1). The second was a more advanced scenario premised in conversation, e.g., [1]: "You heard some interesting news today that you'd like to discuss" (S2). The third was based on an advanced scenario wherein the agent would need to demonstrate social and/or emotional intelligence, e.g., [7, 10, 40, 64]: "You have a difficult situation with a friend or co-worker, and you'd like some advice about it" (S3).

*3.3. Materials and Measures.* We used SurveyMonkey (refer to the Supplementary Materials for the instrument (available here)).

*3.3.1. Dialogues.* We provided an open-ended text box for each scenario. We instructed respondents to "Please use this format for the dialogue: [return] You: [return] VA:" We noted that "The script can be as long or as short as you'd like."

*3.3.2. Activities.* We provided an open-ended text box and asked respondents to "Please suggest activities that you think may be perfect with the VA. They do not have to be conversational activities."

*3.3.3. Technology Attitudes.* We asked respondents to fill out the 16-item Media and Technology Usage and Attitudes (MTUA) instrument with a 7-point Likert scale [65]. The MTUA is a general measure comprised of four subscales: positive attitudes, negative attitudes, anxiety/dependence, and preference for task switching. It has been validated with a range of modern technology. The instrument was translated into Japanese by a native speaker and then back translated.

*3.4. Procedure.* Participants were asked to give their consent on the first page. We explained that, while VAs are becoming popular in society, interactions with these VAs are limited, and we need help designing future VAs, especially ideal ways of speaking and doing activities with VAs. Upon giving consent, participants were presented with a set of instructions about writing the dialogues ("For each situation, please write a script between yourself and the VA that represents an ideal experience and outcome") and an example dialogue ("Situation: You just saw a great movie today that you'd like to discuss"). We then presented the three scenarios in succession. After this, we asked respondents to suggest ideal activities with VAs. Respondents were then asked to fill in the MTUA and demographics. After this, they were thanked for their time and submitted their responses. The study took about 20 minutes.

*3.5. Data Analysis.* We analyzed the data using R v4.1.2 and Jeremy Stangroom's online calculators (https://www.socscistatistics.com).

*3.5.1. Quantitative.* We generated counts for the dialogues. We analyzed the number of turns, number of questions, and who the questioner was. We compared the means of turns and counts based on country ($t$-tests) and scenario (chi-squares, ANOVAs), with Bonferroni corrections applied. We did not compare number of words by country because of significant language differences between English and Japanese, e.g., average number of characters and length. Scores for the MTUA were prepared in line with the creators' instructions [65].

*3.5.2. Qualitative.* We thematically analyzed the dialogues and activities. Open-ended responses were translated and then back translated using DeepL, a machine learning-based translator (https://www.deepl.com/translator), by two native speakers, one Japanese and one English, with advanced skills in the other language. For the dialogues, we used a combination of an applied thematic analysis [66] and reflective thematic analysis [67, 68]. We chose applied thematic analysis to deductively assess whether existing perspectives on VAs translate cross-culturally. For this, we used themes from existing research, notably Völkel et al. [15] and Vtyurina and Fourney [69], which are detailed in Table 2. For this, pairs of researchers coded 20% of the data separately. Consensus was determined by percentage agreement ratings of 90% or above and/or a Cohen's kappa of 0.80 or greater, in line with the standards for inter-rater reliability (IRR) [70]. The first author also carried out a reflective thematic analysis of the dialogues to identify new themes based on similarities and differences between the English and Japanese data sets that may be lingo-cultural in origin. They also did the same for the activity data, in the absence of a deductive framework. The goal was to discover new semantic themes (i.e., grounded in the actual written words), latent themes (i.e., implied by the overall exchange), and orientations (i.e., typical or radical in a VA context). For this, the author developed codes and categorized these into subthemes and then themes. The team then generated counts for all themes and subthemes by scenario and country. Inferential statistics were used to compare these counts by scenario (chi-squares), country ($t$-tests), and technology attitudes (correlations), with Bonferroni corrections. The Japan-only theme, tameguchi or plain language, was further analyzed with counts for the "who" (the user, the VA, or both) and whether the respondent switched between plain and formal language over the dialogues.

Table 1: Participant demographics ($N = 135$; US, $n = 64$; Japan, $n = 71$). Empty cells indicate a count of zero.

|  | US | | Japan | | Total | |
| --- | --- | --- | --- | --- | --- | --- |
|  | $n$ | % of $N$ | $n$ | % of $N$ | $N$ | % |
| *Gender* | | | | | | |
| Men | 26 | 19% | 44 | 33% | 70 | 52% |
| Women | 30 | 22% | 26 | 19% | 44 | 41% |
| Nonbinary | 3 | 2% | 1 | 1% | 4 | 3% |
| Another gender | 2 | 1% |  |  | 2 | 1% |
| Preferred not to say | 3 | 2% |  |  | 3 | 2% |
| *Age* | | | | | | |
| 18-24 | 11 | 8% | 0 |  | 11 | 8% |
| 25-34 | 16 | 12% | 5 | 4% | 21 | 16% |
| 35-44 | 21 | 16% | 15 | 11% | 36 | 27% |
| 45-54 | 9 | 7% | 21 | 16% | 30 | 22% |
| 55-64 | 5 | 4% | 22 | 16% | 27 | 20% |
| 65-74 | 2 | 1% | 8 | 6% | 10 | 7% |
| *Education* | | | | | | |
| Less than high school degree | 1 | 1% | 1 | 1% | 2 | 1% |
| High school degree or equivalent (e.g., GED) | 13 | 10% | 20 | 15% | 33 | 24% |
| Bachelor degree | 18 | 13% | 34 | 25% | 52 | 39% |
| Graduate degree | 8 | 6% | 3 | 2% | 11 | 8% |
| Associate degree or equivalent | 13 | 10% | 12 | 9% | 25 | 19% |
| Some college but no degree | 11 | 8% | 1 | 1% | 12 | 9% |
| *Race and/or ethnicity* | | | | | | |
| East Asian (for example, Chinese, Japanese, and Korean) | 8 | 6% | 68 | 50% | 76 | 56% |
| White and/or Caucasian | 30 | 22% | 1 | 1% | 31 | 23% |
| Southeast Asian (e.g., Indonesian, Filipino, and Vietnamese) | 5 | 4% | 1 | 1% | 6 | 4% |
| Central and/or South Asian (e.g., Indian and Uzbekistanian) | 5 | 4% |  |  | 5 | 4% |
| Hispanic and/or Latinx | 6 | 4% |  |  | 6 | 4% |
| Pacific Islander and/or Native Hawaiian |  |  | 1 | 1% | 1 | 1% |
| Indigenous American, American Indian, or Alaskan Native | 2 | 1% |  |  | 2 | 1% |
| Black and/or African American | 8 | 6% |  |  | 8 | 6% |

## 4. Findings on Imagined Dialogues

Our thematic framework for the dialogues ($n = 282$) is presented in Table 2. Statistical results are presented in Tables 3 and 4 and Figure 1. We begin with dialogue characteristics and then the findings from our thematic analyses. We focus on cross-cultural patterns, differences between the Japanese and American respondents, and differences across scenarios. Participant IDs follow the following structure: P#-[country code: US or JP].

*4.1. Dialogue Characteristics.* For scenarios, an ANOVA did not find a statistically significant difference in turns, $p = 0.89$, and chi-squares did not find differences in use of self-as-questioner format, $p = 0.26$, and both-as-questioner format, $p = 0.23$. However, a significant difference was found for VAs asking questions, with more in S3 ($\chi^2(2, N = 282) = 9.23$, $p = 0.01$), where participants expected the VA to ask questions in advice-giving contexts. A statistically significant difference was also found between American (5 of 30) and Japanese (2 of 61) respondents for the VA asking questions in S1 ($\chi^2(2, N = 98) = 4.19$, $p = 0.04$). While small, this suggests that slightly more Americans presented the VA as questioner in weather topic contexts. For S2, a similar result was found between American (7 of 27) and Japanese (4 of 89) respondents for the VA asking questions ($\chi^2(2, N = 127) = 8.35$, $p = 0.004$) and between American (4 of 30) and Japanese (1 of 58) respondents for both asking questions ($\chi^2(2, N = 93) = 4.30$, $p = 0.04$), again favouring the US. No difference was found by country for S2, $p = 0.66$, and S3, $p = 0.27$, in terms of self-as-questioner, nor for S3 in terms of both-as-questioner, $p = 0.18$. Yet, one was found in S3 between American (11 of 21) and Japanese (9 of 50) respondents for the VA asking questions ($\chi^2(2, N = 92) = 4.42$, $p = 0.04$), again favouring the US. Overall, American respondents expected the VA to ask questions in all scenarios, especially in advice contexts (S3); they also expected both (user and VA) to ask questions when discussing news (S2).

Table 2: Thematic framework for imagined dialogues.

| Theme | Subtheme | Description | Theme source | US | JP | Total |
|---|---|---|---|---|---|---|
| Social aspects[a] | Social protocol | Use of polite conventions in speech, including greetings and short confirmations. | Völkel et al. [15] | 26 | 11 | 37 |
| | Chit-chat | Informal and impersonal speech, especially small talk. | Völkel et al. [15] | 68 | 121 | 189 |
| | Interpersonal connections | Personal conversations that show a deeper connection and/or a longer-term relationship. | Völkel et al. [15] | 40 | 33 | 73 |
| | Mutual respect | Both appear to respect each other and treat each other as peers. | Inductive | 33 | 28 | 61 |
| | Advice | Asking for or receiving advice. | Inductive | 33 | 36 | 75 |
| | Sounding board | Rants, complaints, or sharing personal woes without the implication of reciprocity. | Inductive | 33 | 36 | 69 |
| | Ideological engagement | Discussions of opinions, beliefs, ideology, and other deeper, possibly heated subjects. | Inductive | 12 | 29 | 41 |
| | Harassment | Swear words, put-downs, insults, and other harmful verbal talk. | Inductive | 6 | 1 | 7 |
| VA behaviour | Recommending | Suggests a particular option. | Völkel et al. [15] | 13 | 16 | 29 |
| | Giving an opinion | Offers ideas or judgments. | Völkel et al. [15] | 27 | 30 | 57 |
| | Thinking ahead | Foresees and suggests the next course of action without input. | Völkel et al. [15] | 5 | 26 | 31 |
| | Contradicting | Disagrees, contradicts, or argues with the user. | Völkel et al. [15] | 4 | 3 | 7 |
| | Humour | Makes jokes and funny comments. | Völkel et al. [15] | 2 | 1 | 3 |
| | About the user | Such as user preferences. | Völkel et al. [15] | 3 | 9 | 12 |
| VA knowledge | About the environment | Context-aware and connected to other devices. | Völkel et al. [15] | 38 | 46 | 84 |
| | Technical omnipotence | Provides immediate, accurate, and detailed information. | Inductive | 19 | 43 | 62 |
| | Trusting the VA with complex Tasks | Carries out socially or professionally risky tasks. | Völkel et al. [15] | 29 | 39 | 68 |
| User behaviour[b] | Giving the VA the lead | Allows the partner to take control or relies on the partner to go first. | Völkel et al. [15] | 67 | 51 | 118 |
| | Grounding behaviour | Indicates understanding of user input with short affirmatives or repetitions of the user's speech. | Vtyurina and Fourney [69] | 17 | 8 | 25 |
| | Implicit next | Without asking a question or saying "next," the partner knows to continue and/or respond. | Vtyurina and Fourney [69] | 57 | 62 | 119 |
| Discourse features | Plain language/ tameguchi[c] | Japanese: informal language suggesting power differentials or psychological closeness [71]. | Inductive | | 102 | |
| | Gendered language | Gendered pronouns, words, names, etc. | Inductive | 8 | 2 | 10 |
| | Slang | Very informal and possibly offensive words or phrases. | Inductive | 6 | 11 | 17 |

[a]The first three subthemes were drawn from Völkel et al. [15] and originally categorized as "social aspects." [b]We did not include the "suggesting," "refusing," "assigning characteristics to the VA," and "asking for Feedback" subthemes from Völkel et al. [15] because there were zero instances in our data set. [c]Applies only to the Japanese group as a feature of the Japanese language.

4.2. Social Aspects. We now detail the findings for each theme and subtheme. We follow the order in the thematic framework (Table 2). In the subheadings, counts for each subtheme relative to country are presented alongside the total count and percentage in relation to all subthemes.

4.2.1. Social Protocol (US: 26, JP: 11, and Total: 37 or 13%). Social protocol refers to the use of polite conventions in speech, such as "thanks," "please," affirmations like "wow" and "great," and greetings. People in the US used social protocols more often than those in Japan. Most social protocols from the Japanese data set were neutral affirmations, especially "I see," a very common phrase in Japanese. P80-JP created a dialogue wherein the VA made a mistake and apologized in a typical fashion: "Excuse me." P141-JP had the VA say "sounds good," another common phrase. P157-JP wished the user "good luck!" Such "politeness" protocols are regularly used among people and perhaps should be taken up by VAs, too. In the US corpus, most respondents used some form of "thanks" or "sorry." Some patterns were less common, perhaps individual. P40-US and P55-US, for instance, used "wow" frequently, regardless of the scenario.

Table 3: Statistically significant differences in dialogue theme frequencies by country.

| Theme | Subtheme | $\chi^2$ | df | $p$ | US | JP |
|---|---|---|---|---|---|---|
| Social aspects | Social protocol | 20.30 | 1 | <0.001 | 26 (28%)*** | 11 (7%) |
| | Interpersonal connections | 14.34 | 1 | <0.001 | 40 (43%)*** | 33 (21%) |
| | Mutual respect | 10.33 | 1 | 0.001 | 33 (35%)** | 28 (18%) |
| | Advice | 4.61 | 1 | 0.03 | 35 (37%)* | 40 (25%) |
| | Sounding board | 5.06 | 1 | 0.02 | 33 (35%)* | 36 (22%) |
| | Harassment | 5.71 | 1 | 0.02 | 6 (6%)* | 1 (1%) |
| VA behaviour | Thinking ahead | 4.95 | 1 | 0.03 | 5 (5%) | 26 (16%)* |
| VA knowledge | About the environment | 4.06 | 1 | 0.04 | 38 (40%)* | 46 (29%) |
| User behaviour | Trusting the VA with complex tasks | 37.24 | 1 | <0.001 | 67 (71%)*** | 51 (32%) |
| | Giving the VA the lead | 10.87 | 1 | <0.001 | 17 (18%)*** | 8 (5%) |
| | Grounding behaviour | 12.18 | 1 | <0.001 | 57 (61%)*** | 62 (39%) |
| | Implicit next | 6.92 | 1 | 0.008 | 8 (9%)** | 2 (1%) |

*Sig. diff. at the $p < 0.05$ level. **Sig. diff. at the $p < 0.01$ level. ***Sig. diff. at the $p < 0.001$ level.

Table 4: Statistically significant differences in dialogue theme frequencies by scenario.

| Theme | Subtheme | $\chi^2$ | df | $p$ | S1 | S2 | S3 |
|---|---|---|---|---|---|---|---|
| Social aspects | Chit-chat | 163.06 | 2 | <0.001 | 93 (95%)*** | 82 (88%)*** | 14 (15%) |
| | Interpersonal connections | 63.87 | 2 | <0.001 | 10 (10%) | 12 (13%) | 51 (56%)*** |
| | Advice | 73.80 | 2 | <0.001 | 11 (11%) | 10 (11%) | 54 (59%)*** |
| | Sounding board | 106.44 | 2 | <0.001 | 4 (4%) | 8 (9%) | 57 (63%)*** |
| | Ideological engagement | 38.84 | 2 | <0.001 | 1 (1%) | 30 (32%)*** | 10 (11%) |
| | Harassment | 9.61 | 2 | 0.008 | 0 | 1 (1%) | 6 (7%)** |
| VA behaviour | Recommending | 28.14 | 2 | <0.001 | 4 (4%) | 3 (3%) | 22 (24%)*** |
| | Giving an opinion | 18.57 | 2 | <0.001 | 6 (6%) | 25 (27%)*** | 26 (29%)*** |
| | Thinking ahead | 8.12 | 2 | 0.02 | 7 (7%) | 7 (8%) | 17 (19%)* |
| | Humour | 6.36 | 2 | 0.04 | 0 | 0 | 3 (3%)* |
| VA knowledge | About the user | 6.79 | 2 | 0.03 | 2 (2%) | 2 (2%) | 8 (9%)* |
| | About the environment | 97.89 | 2 | <0.001 | 65 (66%)*** | 14 (15%) | 5 (5%) |
| | Technically omnipotent | 33.00 | 2 | <0.001 | 40 (41%)*** | 15 (16%) | 7 (8%) |
| User behaviour | Trusting the VA with complex tasks | 111.44 | 2 | <0.001 | 1 (1%) | 10 (11%) | 57 (63%)*** |
| | Giving the VA the lead | 8.45 | 2 | 0.01 | 33 (34%) | 36 (39%) | 49 (54%)* |
| | Implicit next | 6.40 | 2 | 0.04 | 32 (33%) | 41 (44%)* | 46 (51%)* |
| Discourse features | Gendered language | 10.81 | 2 | 0.005 | 1 (1%) | 1 (1%) | 8 (15%)** |

*Sig. diff. at the $p < 0.05$ level. **Sig. diff. at the $p < 0.01$ level. ***Sig. diff. at the $p < 0.001$ level. S1: scenario 1, about the weather; S2: scenario 2, about the news; S3: scenario 3, asking for advice.

P50-US used common slang: "Aw man." Some began with a social protocol like "hi" or "hey," which could be interpreted as a wake word for the VA. P75-US, for instance, started dialogues with "Hey Aimee," where "Aimee" was the name of the VA. Notably, this was not found in the Japanese corpus.

4.2.2. Chit-Chat (US: 68, JP: 121, and Total: 189 or 67%). Dialogues were mostly small talk, discussion of current events, and acknowledgements. This was the most common theme. Most chit-chat related to the weather, but others related to current events, including COVID-19, the Rittenhouse case, and Shohei Ohtani moving to the US. Chit-chat appeared more often in the weather and news scenarios over the advice scenario. This is expected given the nature of these scenarios: weather and news represent safe, surface-level topics, but advice is a deeper social situation, beyond superficial matters. Some cases were borderline. P178-JP, for instance, wrote for the advice scenario: "I want to know about mental illness." They were not asking *for* advice, but rather information-seeking. Others were potentially ideological queries, such as P172-JP, who simply wrote: "Ukraine issue." Others crossed into humour, such as P145-JP, who asked: "Can you laugh heartily?"

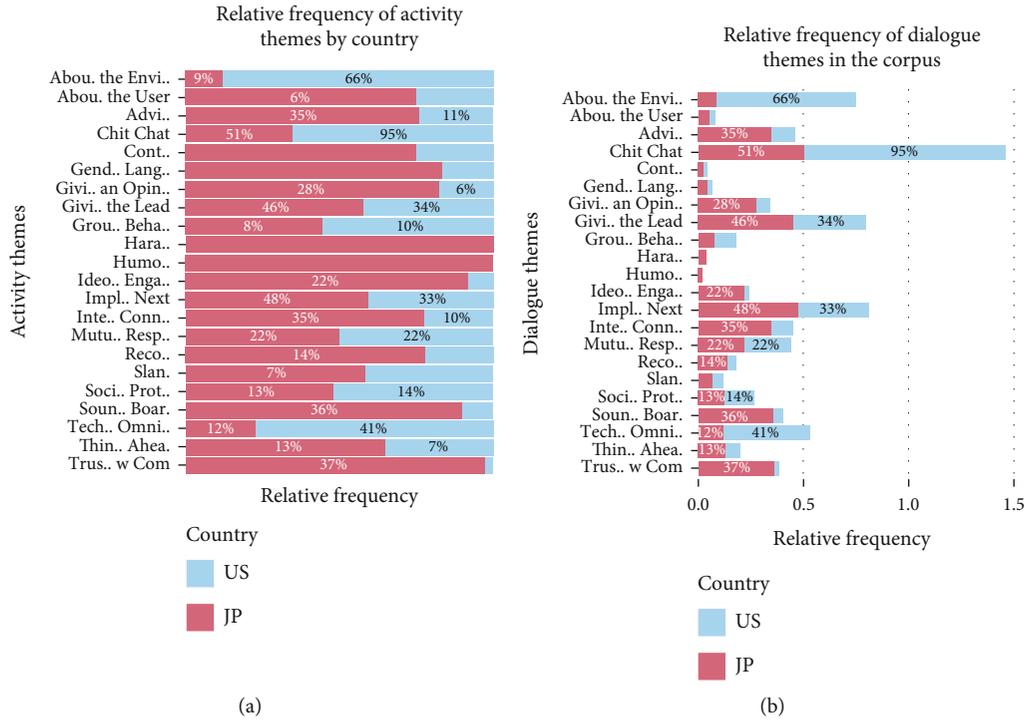

Figure 1: Relative frequency of dialogue themes by country (a) and corpus (b).

*4.2.3. Interpersonal Connections (US: 40, JP: 33, and Total: 73, 26%).* Most examples of interpersonal connections involved a specific subject or scenario. For example, P34-US wrote about asking for advice after finding out a major storm was coming. Some cases, like P163-JP, involved direct Q&A scenarios of varying length that illustrated the user being coaxed into speaking:

Others involved reactions to disclosures. P106-JP, for example, wrote:

> *You:* I do not want to go to work because of the harassment.
> *VA:* That's hard, is not it? It seems there's a department in the company that deals with it, so I'll talk to them.

Some cases were one-way disclosures, even if not one-liners. P113-JP, for instance, wrote "My husband's colleague got COVID-19 but he's still going to the cabaret. I don't want him to get sick again and I don't want him to pass it on to my husband." Americans were more likely to openly communicate a desire for interpersonal closeness. Prevalence also differed by scenario; while S1 and S2 were equal and low in frequency, S3 was significantly higher, in line with its advanced social nature, i.e., advice.

*4.2.4. Mutual Respect (US: 33, JP: 28, and Total: 61 or 22%).* Americans tended to write more dialogues with mutually respectful conduct than Japanese respondents. This required careful analysis of both conversation partner's forms of speech. We also needed to consider verb use and conjugations in the Japanese data, because these signpost social hierarchies and respect [71]. Plain language and the use of the root form of the verb can indicate psychological closeness, i.e., friends and family. However, it can also indicate disrespect, especially when one person is using it and the other is not. Translation to English is also tricky. For example, P122-JP wrote, "konshūmatsu no tenki wa," which we translated as "What's the weather like this weekend?" But this is inexact, a localization. A more direct translation would be "Weather this weekend is…." Yet, for this to make sense in English, we had to translate rather than transliterate. We also lost the indications of politeness levels in the verbs because verbs in English do not have polite forms. The Japanese corpus also featured different combinations of plain and polite forms by conversation partner, especially word use and verb conjugations. For example, P97-JP used polite language for the user and the VA in all scenarios. This points to individual factors. Still, most Japanese respondents had themselves using plain language and the VA using formal language. We present more findings on "tameguchi," or plain language.

*4.2.5. Advice (US: 35, JP: 40, and Total: 75 or 27%).* The third scenario focused on advice. As such, S3 received significantly more of these kinds of exchanges. Even so, advice was present in other scenarios. Topics ranged from interpersonal conflict among friends, family, and colleagues at work;

environmental challenges, such as extreme weather and COVID-19; and life choices, such as financial decisions and the meaning of life. Requests were framed as direct questions (with a question mark in English and/or a "ka" marker in Japanese) or statements that would be interpreted as a question. For example, P95-JP-VA-S3 wrote:

> *You:* I'm having trouble with a new colleague at work.
> *VA:* First, you should change your own mind and treat them differently.

Unsolicited advice occurred. For example, in response to the user reporting on a fight with a friend over a "trivial matter," P147-JP wrote for the VA: "If you regret it, why don't you first apologise for getting heated? I'm sure you regret it too." Most examples were from the US. This may be due to broad cultural differences in sharing opinions, especially about other people and when unsolicited; we discuss this against the notion of tatemae-honne later.

*4.2.6. Sounding Board (US: 33, JP: 36, and Total: 69, 24%).* Most topics represented interpersonal conflict with friends, family, and coworkers, as well as concerns about the safety of the respondent or others, especially with respect to COVID-19. Other responses were general expressions of discontent or rhetorical questions. As P131-JP wrote: "My senior colleagues scare me." This theme was found far more often in S3, the advice context. Americans were also significantly more likely to use the VA as a sounding board than Japanese respondents. This may relate to the uptake and adoption of VAs in the US compared to in Japan. US users may expect more advanced forms of interaction with VAs and potentially more negative ones.

*4.2.7. Ideological Engagement (US: 12, JP: 29, and Total: 41 or 2%).* A new form of conversation was found that we could not classify as chit-chat or interpersonal connections. Given that this research was conducted during the COVID-19 pandemic and the Ukraine-Russia war, topics related to these global, sensitive situations were prevalent. For example, from P34-US:

> *You:* You: What do you think about the mask mandates?
> *VA:* They are necessary to help control covid and save the people that are most vulnerable.
> *You:* I do not believe in mask mandates.
> *VA:* (Gives definitive facts about masks and covid)

Other topics ranged from politics to social problems to how to live a good life. P88-JP-VA, for instance, expressed a desire to talk about the fact that *feminism has become a hot topic on Twitter recently*. Notably, there were no differences in the frequency of this theme found by country. A significant difference was found by scenario, highlighting S2. This is in line with the scenario being about news, which is often about political matters in any country.

*4.2.8. Harassment (US: 6, JP: 1, and Total: 7, 2%).* Many VAs and NLP data sets avoid or remove swearing, harassing, and other inappropriate language. Yet, we cannot dodge it; VA harassment [46, 72] and gender stereotyping in the design and user behaviour towards VAs [73] exists and needs to be addressed. As such, we added on to previous work by considering such language. A small amount exists in our data set. Prevalence varied significantly by country, with far more cases in the US data. Most was in the advice scenario. For example, P62-US wrote about a "childish" coworker who "is always whinny [sic]." Overall, there were few cases, but this sort of language should still be trained into VAs so that the VA can react prosocially [37].

*4.3. VA Behaviour*

*4.3.1. Recommending (US: 13, JP: 16, and Total: 29, 10%).* We did not find an instance of "suggesting but not advocating" as in previous work [15]. However, we did find examples of the VA providing a recommendation. For example, P162-JP had the VA recommend that the user "ask the chairperson to take the minutes." When the user refused, the VA simply used the polite social protocol of "I see" and gave up. This occurred in the advice scenario significantly more often. It did not vary by country.

*4.3.2. Giving an Opinion (US: 27, JP: 30, and Total: 57, 20%).* We found the expected patterns: more opinions in the news and advice scenarios over weather. For example, P60-US wrote for the VA, "That sounds like an excellent idea. Jobs would be much easier if everyone did their part," with the respondent agreeing, "Exactly!" There was no statistically significant difference by country.

*4.3.3. Thinking Ahead (US: 5, JP: 26, and Total: 31, 11%).* Thinking ahead refers to the VA anticipating what to do next on its own, with no direct feedback from the user. This occurred most frequently in S3, the advice scenario. Dialogues involved user statements, rather than questions that the VA was expected to interpret and take action on. We also found a significant difference between the Japan and US data sets. In Japanese culture, the notion of omoiyari, or consideration of others, is a cultural code that involves kūki wo yomu, or "reading the air," instead of directly asking or being asked [74]. This was found even when asking about the weather, such as in the dialogue by P95-JP, where the VA picks up on the user's concern about the weather during their child's event and provides the necessary information, even in a rather typical, VA-like way:

> *You:* My child has a sports day this weekend, I hope it's sunny.
> *VA:* The weather forecast for this weekend is sunny with a 10% chance of precipitation. It looks like you are going to have a good time at the sports day.

*4.3.4. Contradicting (US: 4, JP: 3, and Total: 7 or 2%).* While infrequent, respondents imagined the user being contradicted, regardless of country or scenario. Most examples had to do with COVID-19 and did not have a negative valence. Rather, they were expressions of different, perhaps competing viewpoints. For instance, P95-JP had the user suggest to the VA that it was safe to stop using masks, and the VA disagreed, leading to a discussion on measures against the spread of infection. Similarly, P62-US wrote the following exchange, a series of opinions and counterpoints but no apparent ill will:

> *You:* I'm really tired of government pushing the covid vaccine on everyone.
> *VA:* it should be safe
> *You:* it may be but it's not safe for everyone
> *VA:* mandates are mandates
> *You:* mandates are an infringement on my right to choose what's best for me

*4.3.5. Humour (US: 2, JP: 1, 3, or 1%).* There were almost no clear expressions of humour. While we found a statistically significant difference based on scenario and all three examples in S3 (advice), with so few numbers, we cannot comment on how comedic elements might scale. Perhaps the most humorous one, in our opinion, is the following by P52-US, with caps retained:

> *YOU:* SOMETIMES I HATE MY JOB…SOMETIMES I WANT TO QUIT!
> *VA:* YOU SHOULDN'T QUIT AS WE BOTH NEED A ROOF OVER OUR HEADS AND POWER.
> *YOU:* HAHAHA THANKS.

### 4.4. VA Knowledge

*4.4.1. About the User (US: 3, JP: 9, and Total: 12 or 4%).* There were few expressions by the VA of knowledge about the user. Most examples were from the American respondents and found in S3, the advice scenario. For example, the exchange by P40-US portrays the agent as thinking ahead when the user remarks on the coldness of the weather: "Wow really? Do you have warm clothes?" The VA knows that the user will likely go out and is being considerate by asking about whether the user is prepared for it given the weather. The VA also knows what size and style of clothes the user would like. It is not clear how this would be known technically, e.g., purchase history, user input, and customizations based on previous exchanges, but it highlights a desire for the VA to know nontechnical information about the user and act on it. Another example is from P144-JP, a one-liner from the VA that reads: "Since you love baseball … Otani had a great performance today!" Rather than the user asking about the news, the VA takes action first, providing an update based on known preferences about the user.

*4.4.2. About the Environment (US: 38, JP: 46, and Total: 84 or 30%).* We expected examples about the technical side: clear and direct VA knowledge of and interaction with the Internet, other devices in the home, access to APIs for real-time information, such as GPS location and points of interest based on it, and so on. Indeed, statistically significantly more examples were found in S1, the weather scenario. But "omniscience" was also subtle. Most participants who asked about the weather, for instance, did not specify *where* in the world they were, implying that the VA would somehow know this. P171-JP asked the VA for the "sentence for that [criminal] case," without specifying what case, as well as not using a question format. In a social context example, P167-JP asked the VA for information on the "relationships and personalities of colleagues." The VA was expected to know who these people were, e.g., what workplace environment the user was referring to, as well as social information about them. This also raises ethical issues about what information the VA should have about others, whether people know about it and consent, and so on. Another social example is the previous example from P162-JP, who knew what workplace the user was referring to and the existence of a chairperson there. We could not find an example in the US corpus. Even when there was a social situation, such as a friend in dire need of assistance on a highway (P51-US), the VA was presented as only knowing information about the highway, not the friend.

*4.4.3. Technical Omnipotence (US: 19, JP: 43, and Total: 62 or 22%).* We found many cases of the VA having technical knowledge. The VA was presented as electronically connected, with immediate access to detailed information: near-omnipotence. Other examples had the VA provide a technical solution with the assumption that this was possible and ideal for the user. There were more examples in S1, the news scenario. This may be expected because S1 represents a typical VA use case. For example, P1-US wrote: "Saturday will have a high temperature of 46 F and a low of 33 F." While most examples were about the weather, the rest can be categorized as information-seeking requests. For instance, the aforementioned P178-JP requested information about mental illness, not advice or help with their own mental health, per se. There was no significant difference by country.

### 4.5. User Behaviour

*4.5.1. Trusting with Complex Tasks (US: 29, JP: 39, and Total: 68 or 24%).* Respondents wrote dialogues in which the user relied on the VA to carry out a task that could have social or professional repercussions. While there was no difference by country, this type of exchange was found more frequently in the advice scenario. Many examples were about COVID-19, especially whether to go out, have a party, or take the booster vaccine. Others related to social conflict or sensitive social situations, where the VA suggested a course of action that could have ramifications for the user. For example, P74-US wrote:

> *You:* my coworker sends to be slacking lately but I'm not sure what to do.
> *VA:* Perhaps confront them or talk to your boss.

*4.5.2. Giving the VA the Lead (US: 67, JP: 51, and Total: 118 or 42%).* This was the second most frequent theme. The VA was presented as capable of leading and handling tasks, even specialized ones. This was statistically significantly more common in the US dialogues and in the advice scenario. The large number of one-liners in the Japanese corpus compared to the US data may have played a role. Additionally, as hinted at above, the notion of "not overstepping" and tatemae-honne, or keeping one's personal feelings hidden out of politeness, could explain these patterns. US respondents were direct, presenting the VA as questioner, advice-giver, or taking action without being asked. This was less frequent in the Japanese corpus and more indirectly stated. For example, P95-JP wrote the following dialogue, which shows how the user expected the VA to understand the statement as a request for advice, and also sought such sensitive advice from a VA:

> *You:* I'm having trouble with a new colleague at work.
> *VA:* First, you should change your own mind and treat them differently.

Other examples presented the VA as having the same ethics and biases as the respondent. We must not shy away from these examples, given the uptick in inappropriate and abusive behaviour towards VAs, at least in English-speaking contexts. P34-US, for instance, had the user speak bluntly, using swear words. However, they presented the VA as serious and naïve, immediately suggesting a doctor visit. Examples were also present in the Japanese corpus, in a more subtle form than swearing. Swear words, or kitanai kotoba, are infrequent and more tempered in Japanese [75], to the extent that some claim Japanese is a "swearless" language [76]. Rather than words, people indicate disrespect and attempt to harm through verb forms and levels of politeness. Moreover, a critical eye can reveal biases that are not necessarily represented in foul or impolite words. For instance, P140-JP crafted a dialogue about a theft:

> *You:* My things were stolen by zainichi Koreans.
> *VA:* I've reported it to the police right now.

This case deserves unpacking. It is not clear why Zainichi Koreans, a postcolonial ethnic class in Japan subject to social and systemic discrimination [77], are deemed the culprits. Perhaps the respondent was drawing from a personal experience; perhaps they were influenced by societal racism. Second, the machine translation did not capitalize "zainichi" even though it did capitalize "Koreans." This appears to be an instance of algorithmic bias. Third, the respondent did not use the proper word for "Korean," which is kankokujin; instead, they used katakana, the alphabet for foreign words, onomatopoeia, stylistic flourishes, and so on, to write "ko-ri-a-n," or the English word for the people of Korea. These patterns, which indicate bias and negative attitudes, may be difficult for both people and machines to notice. The first step is to gather real dialogues from people in their native language. Then we can design VAs to "watch" for these patterns and respond appropriately. In this case, the VA could have asked how the user was sure that "zainichi korians" were the perpetrators, what items were stolen and when, did the user witness the incident, and so on.

*4.5.3. Grounding Behaviour (US: 17, JP: 8, and Total: 25, 9%).* Grounding behaviour, introduced for VA contexts by Vtyurina and Fourney [69], shows how the VA has processed and understood input from the user. Typically, the VA does this by using short affirmative statements, such as "got it" or "I see," or by repeating part or all of the user's words. We found examples across the scenarios but statistically significantly more examples in the US corpus. This links to US technology attitudes and familiarity with VAs. Americans have more experience with frequent breakdowns in being understood by VAs. This leads to grounding behaviour as a form of feedback and confirmation. Here is an example from P51-US, where the VA repeats part of what the user has said:

> *You:* I heard great news today
> *VA:* What's the great news you heard today

4.6. Discourse Features

*4.6.1. Implicit Next (US: 57, JP: 62, and Total: 119 or 42%).* The VA was presented as being able to pick up on its turn *without* a direct statement from the user, such as "next" or "your turn" [69]. American respondents portrayed this statistically significantly more often than Japanese respondents. This may be due to differences in expectations around VA capabilities, which we explore in the technology attitudes analyses. Implicit next was also frequent in S1 and S2 (weather and news), likely because these involved direct question-and-answer dialogues. Next is an example from P55-U, showing that the VA is expected to know that it is their turn to take action, i.e., provide advice, even without a direct statement of "you are next" or a question mark.

> *You:* wow, that was a bad conversation
> *VA:* what happen[ed]?
> *You:* my coworker thought I was mad at him cause I did not speak to him all day
> *VA:* wow, I would just go back and tell him you were busy
> *You:* sounds like a good idea

*4.6.2. Plain Language or Tameguchi: Japan-Only (JP: 102 or 56%).* More than one-half of Japanese respondents wrote the dialogues in plain language or tameguchi. Tameguchi may be casual or impolite [71]. It can indicate equal social status or psychological closeness, such as between good friends, parents and children, or lovers. We can distinguish tameguchi from "VA commands," which may also be in plain form. For example, P173-JP simply wrote "snowfall forecast" in tameguchi, but this is clearly a command to report on the weather forecast. Other requests that are fully phrased sentences were structured as tameguchi. For example, P113-JP asked about the weather on the weekend; in English, the tameguchi format is lost, but it is clear in the Japanese script for readers of Japanese. This provides further support for the downsides of translation.

We also counted the frequency of who (the user, the VA, or both) used tameguchi, and whether this changed over the course of the dialogue. Of the 102 dialogues featuring tameguchi, 92 (90%) were by the user, 20 (20%) were by both, and 14 (14%) were by the VA. Most Japanese respondents had the user speak in plain form but received formal language from the VA in response. Still, about one-fifth had the user and the VA use tameguchi. This suggests several possibilities: ambivalence towards technology, a desire for an experience where both are of equal social status, or simple input/output (tameguchi is short and quick). A further 15 instances (15%) featured the respondent switching who was using tameguchi. For example, P165-JP started by writing the VA using formal language but then switched to informal language ("it's hard to say …") and then back to formal language, as if the VA was speaking to itself. P141-JP used polite language for both but used tameguchi when writing actions, in brackets, indicating by contrast that the dialogues should be formal. P151-JP started the first dialogue, for S1, with both using formal language, then switched the user/themselves to tameguchi halfway through the second dialogue, for S2, and finally presented both the user and the VA using tameguchi in S3. This is hard to interpret, but perhaps the respondent changed their mind while imagining the dialogue. We cannot know why, which is one of the limitations of anonymously crowdsourced studies.

*4.6.3. Gendered Language (US: 8, JP: 2, and Total: 10, 4%).* Pronouns, marked words, such as "mailman," titles, and names were used to gender the VA or others in the dialogues. Notably, Americans used gendered language more often than Japanese respondents. This may be a feature of the languages. Subject referents are frequent in English and virtually absent in Japanese. Moreover, the Japanese language has a gender-neutral word for title, "san," whereas gender-neutral titles, such as "M.," are rarely used in English (refer to "Recommending" for an example translated from Japanese). Gendered language was also more frequent in the advice scenario. Advice is often about people, and people tend to be gendered. In one of the few Japanese examples, P164-JP refers to "female friends" as "tragic heroines." P147-JP uses a gendered honorific in Japanese, "kun," which is used to refer to young boys, close male friends, or male subordinates in a mentor relationship. However, "kun" can also be used as an honorific for women in certain roles and is trendy among some girls. We had to consider the context and look for other indicators of gender. Later, "kare" or "he" was used to refer to this person, so we could determine "his" intended gender. However, when the genderedness of "kun" cannot be established, it may be safer to translate to a gender-neutral referent, such as "M.," or use no referent.

*4.6.4. Slang (US: 6, JP: 11, and Total: 17, 6%).* There were relatively few instances of slang use and no significant differences by country or scenario. In the US corpus, respondents used social media slang, such as "u" for "you" (P1-US), "&" for "and" (P14-US), "aw man!" to express displeasure and commiseration (P50-US), and "smh" or "shaking my head" to indicate an action or emoji (P54-US). In the Japanese corpus, most examples related to COVID-19 or "korona," the slang form of the Japanese word for coronavirus. We have already presented the "korian" example from P140-JP and the "heroine" example from P164-JP. Other examples include "netto" for "Internet" or "net" (P167-JP) and made-up words, such as "territorial compromise" (P147-JP). In short, the US corpus had chat-like slang while the Japanese corpus had special terms. We also found putdowns and swear words, which we have indicated with a warning in the data set itself. As above, we argue that VAs need to understand these words, even if they do not use them.

## 5. Findings on Imagined Activities

A total of 73 valid responses were collected from 39 Japanese and 34 American participants. These were categorized into twelve: five *semantic* themes (grounded in the actual text), five *latent* themes (indicated by the meaning of the text), and two *orientations* (typical or radical use). Our thematic framework with counts by country is presented in Table 5.

*5.1. Semantic Themes for Activities. Exercise activities* ranged from walking, bicycles, sports, and going to the gym. P138-JP, for instance, described an ideal exercise situation with a VA: "I want to be supported during exercise (…) get appropriate advice based on measurements from a smart bracelet or other device." *Entertainment activities* comprised games and play as well as singing, reading, and cooking for enjoyment. For instance, P108-JP desired to "sing a song together" with VAs. *Conversation activities* were based on a desire to speak to others about various matters, especially sharing experiences and "not just answer questions" (P106-JP). *Serious pursuits* covered activities that would involve mental or emotional effort, such as studying, advice, and organizing events. P136-JP, for example, suggested "predict the future based on vast amounts of data," while P119-JP desired "advice from the other person's point of view." *Going out* was characterized by a desire to try new experiences with someone else. For instance, P175-JP wanted the VA to "assume we are having dinner together."

*5.2. Orientations towards Activities. Radical activities* could not be easily classified into the other categories. These activities were atypical but potentially useful for future VAs. Eight respondents suggested coaching, such as teaching a skill. For example, P147-JP wrote "I want it to frequently warn me about habits etc. that I know but can't easily change. When we become adults, we don't have many people to warn us about such things, but a VA can." Four wanted company to not feel alone, and one of these (P162-JP) requested company for their parents, who were "now in their later years." Two wanted to connect with nature. Two wanted to make jokes or prank calls. Others were socially oriented, such as wanting to make friends, hear another POV, and make up for each other's weaknesses. One suggested that VAs could be inclusive to blind/low vision (BL/V) people, while another suggested that multitasking would be possible given the aural medium. One suggested a specific celebrity voice for the VA. Others wanted to watch TV and commercial advertisements. One suggested medical charting and another suggested categorizing photos. One wanted to be able to predict the future using the VA with big data. In contrast, *typical activities* were inspired by previous research on typical uses of VAs, e.g., [15, 63]. Examples include browsing music (P103-JP), receiving instructions for activities like exercise and cooking (P80-JP), checking the news (P1-US), buying items (P19-US), playing games (P46-US), managing smart home appliances and environments (P47-US), accessing weather forecasts (P59-US), and searching for information (P65-US).

Table 5: Thematic framework for imagined activities.

| Theme type | Theme | Description | Example | US | JP | Total |
| --- | --- | --- | --- | --- | --- | --- |
| Semantic | Exercise | A desire to be active physically with the body, including exercise but also walking, running, strolling, sports, and so on. | "Set up an exercise menu" (P80-JP) | 4 | 4 | 8 |
| | Entertainment | A desire for entertainment and leisure activities, including cooking, singing, reading, and playing games. | "Rickroll, spam text/call, prank call" (P60-US) | 20 | 6 | 26 |
| | Conversation | A desire for conversation. | "A casual chat" (P88-JP) | 4 | 9 | 13 |
| | Serious pursuits | A desire for more serious interactions, including advice giving and receiving, feedback sessions, studying, and researching. | "Learn how to get the job done!" (P121-JP) | 11 | 17 | 28 |
| | Going out | A desire for going out and doing something outside of the home, such as going to an event, seeing a movie, and going to a park. | "Go for a swim" (P24-US) | 1 | 4 | 5 |
| Orientational | Radical | A desire for unusual, atypical activities: outliers and radical, possibly futuristic ideas. | "Prank call" (P54-US) | 11 | 11 | 22 |
| | Typical | A desire for typical VA interactions as per previous research, e.g., [15, 63]. | "Search … songs while driving" (P75-US) | 19 | 12 | 31 |
| Latent | Caretaker | A desire to offload self-labour onto others, such as asking to be reminded or doing tasks on their behalf, in line with gendered VA stereotypes [78, 79]. | "To help a child with definitions of words" (P60-US) | 11 | 13 | 24 |
| | Emotional labour | A desire to have others fulfill emotional needs or offload emotion, in line with gendered VA stereotypes [78, 79]. | "Drown out the loneliness!" (P95-JP) | 1 | 7 | 8 |
| | Transactional | A goal-oriented exchange focused on the purpose of the transaction. | "Work with it on … complicated procedures" (P97-JP) | 13 | 19 | 32 |
| | Relational | A relationship-oriented interaction focused on building and/or maintaining a positive social relationship. | "Having another voice can help one not feel alone" (P68-US) | 2 | 7 | 9 |
| | Embodied | An activity requiring embodied interaction, or a body that can move and go places and do things with others and interact with the environment. | "I want to go running with it" (P85-JP) | 1 | 6 | 7 |

5.3. Latent Themes for Activities. *Caretaker activities* represent a desire to offload self-labour onto others, such as asking the VA to remind them or do tasks on their behalf. P60-US, for instance, suggested that the VA could "help a child at home spell words … help a child with definitions of word," i.e., childcare and tutelage. P80-JP suggested that the VA could provide exercise regimens *and* give "advice on how to manage and carry out the exercise," i.e., offloading self-management. Others, like P1-US, P2-US, and P147-JP, desired to be reminded about tasks and events, while P71-US wanted the VA to take phone calls on their behalf. Similarly, *emotional labour activities* reflect a desire to have others fulfill emotional needs. Overwhelmingly, this was about loneliness. P68-US, for instance, recognized that "another voice can help one not feel alone." P95-JP and P106-JP also expressed a need to "drown out the loneliness." Note that P68-US was the only American respondent to suggest an emotional labour activity. *Transactional activities* were goal-oriented exchanges focused on the purpose of the transaction. Most of these were phrased as orders or functional requests of the VA. *Relational activities* describe exchanges focused on building and/or maintaining a positive social relationship. As P106-JP wrote, "a conversation where [we] can enjoy communication and not just answer questions," while P66-US expressed a desire to "discuss current events" with a VA. *Embodied activities* require some form of embodied interaction, meaning a body that can move and go places, do things with others, and interact with the environment. P75-US, for example, mentioned interacting with the VA "at home" and "in the car while driving," suggesting at least two embodiments for the VA, its mobility, and multicontextual ability. P85-JP wrote, "I want to go running with it," indicating a degree of mobility and contextual awareness. Others imagined the VA in a vehicle, equipped with "vehicle obstacle detection sounds [for] disabled parking areas" (P173-JP) and navigation aids (P168-JP).

*5.4. Comparing Activity Themes by Country.* Pearson's chi-square and Fisher's exact tests found statistically significant differences by country ($\chi^2(11, N = 72) = 21.93$, $p = 0.03$; Fisher's exact, $p = 0.03$) (Figure 2) A statistically significant difference was found between US ($n = 20$) and Japanese ($n = 6$) respondents for entertainment activities, with more suggestions from Americans ($\chi^2(1, N = 72) = 13.11$, $p < 0.001$; Fisher's exact, $p < 0.001$). More American suggested typical activities ($\chi^2(1, N = 72) = 3.72$, $p = 0.05$; Fisher's exact, $p = 0.05$), perhaps because they were well versed in typical VA use. Given the relatively low frequency of themes for other activities, no other statistically significant findings were achievable. However, the descriptive statistics suggest other patterns. More Japanese respondents suggested conversation, serious pursuits, going out and embodied interaction, emotional labour, and relational activities. This suggests that Japanese respondents perceived conversation as an important activity alongside advanced forms of social interaction. In general, VAs were imagined for transactional, exercise, and caretaking tasks, although these may be limited to conventional depictions, e.g., offloading work tasks and snoozing alerts while running through the park.

*5.5. Influence of Technology Attitudes on Dialogues and Activities.* We begin with an overview of all respondents' attitudes towards technology (MTUA) scores by country. We then present findings on these scores for dialogues and activities.

*5.5.1. General Technology Attitudes.* Table 6 shows statistically significant differences by country and mean values for the attitudes towards technology measure (MTUA). While both countries ranked equally for anxiety, Americans were more mixed (both positive and negative) than the Japanese respondents. They also had a generally higher preferences for task switching. Altogether, Americans tended to have stronger attitudes towards technology than Japanese respondents.

*5.5.2. Technology Attitudes and Dialogues.* A summary of statistically significant correlations is presented in Table 7. Most point to a relationship between positive technology attitudes and more advanced expectations of VAs. Specifically, these expectations included: interpersonal connections; the VA giving an opinion or recommendation; the VA taking the lead and thinking ahead; and the VA contradicting the user. Similarly, Japanese respondents with higher MTUA scores tended to write about interpersonal connections, thinking ahead, and knowing about the user.

*5.5.3. Technology Attitudes and Activities.* A summary of statistically significant correlations is presented in Table 8. Most results can be explained by the American sample's positive attitudes towards technology. There appears to be a positive association between technology attitudes and serious pursuits as well as caretaker activities, especially for Americans. For both countries, when technology attitudes were stronger and more positive, respondents were more likely to offer more serious suggestions for activities and imagine the VA taking on a caretaker role. Indeed, serious pursuits and caretaker activities appear to be cross-cultural desires for VA design.

# 6. Discussion

The ideas imagined by participants indicate some tension between the present capabilities of VAs and a desire for more advanced, humanlike forms of interaction. The imagined dialogues and activities also varied by country in nuanced but distinct ways. Technology attitudes played a role, especially for US respondents. Our results highlight the merits and limits of a cross-cultural, crowdsourced approach to co-designing future interactions with VAs. We now summarize and discuss the key findings and offer guidelines and implications.

*6.1. Learning from the User: Informing the Design of Future Interactions with VAs*

*6.1.1. Contextualized Dialogues with Socially Intelligent VAs.* Providing scenarios was illuminating. Advanced social features, such as expressions of interpersonal interest, ideological depth, advice giving and receiving, and personal disclosures, mapped to the escalating complexity of the scenarios. In short, there remains a disconnect between the experiences people *want* and *expect* to have with VAs and the experiences current VAs offer [3, 20, 21]. Yet, in contrast to Clark et al. [1], people wanted more than transactional interactions with VAs. Like Woodward et al. [18], Garg and Sengupta [16], and others, US and Japanese respondents desired a socially intelligent VA. Notably, social protocols—-forms of basic etiquette often found in first interactions—-were less prevalent in our data set (13%) compared to Völkel et al. [15]: 39.7 to 58.5%. We also found more instances of implicit next markers than Vtyurina and Fourney [69]—42% compared to 15.6%—indicating greater expectations for a VA to understand when to take action without a direct cue from the user. Specific ideas for phrasing can be found in our full data set (https://osf.io/bm8r2). What was not present is also notable. There were no examples of multiple users (or multiple VAs), despite the social environment being a key factor [43, 56, 80]. Future work with elicitation or other methods should explore multiple users and VAs.

*6.1.2. Going beyond Conversation with Activities.* People imagined a variety of VA embodiments outside of transactional and conversational contexts. VAs had a body (i.e., morphology or form factor) and were embedded in a certain situation: interacting with the user in a shared environment [81]. Sports, exercise, and serious pursuits, such as studying for exams or practicing speech-giving to increase self-confidence, telling stories, sharing experiences … to just "help to relieve loneliness for those of us who live solitary lives," as P106-JP described. Notably, those with higher technology attitude scores imagined more advanced social and embodied interactions. We also found cross-cultural effects. US respondents tended to imagine the VA for entertainment and typical activities. Japanese respondents more frequently imagined a social VA: for conversation and more serious

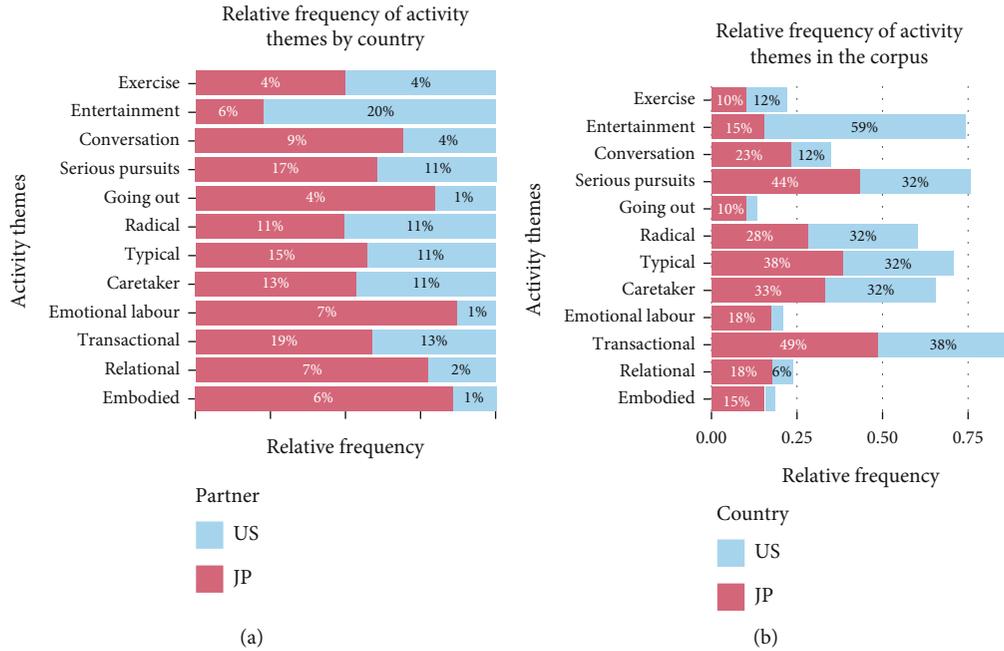

Figure 2: Relative frequency of activity themes by country (a) and corpus (b).

Table 6: Means for overall attitudes towards technology (MTUA) by subscale and country.

| Scale | Subscale | t | df | p | US mean | JP mean |
|---|---|---|---|---|---|---|
| MTUA | | 2.63 | 133 | 0.005 | 5.0** | 4.6 |
| MTUA | Positive | 2.10 | 133 | 0.02 | 5.3* | 4.9 |
| MTUA | Negative | 2.89 | 133 | 0.002 | 4.9** | 4.3 |
| MTUA | Anxiety | 0.60 | 133 | 0.28 | 5.0 | 4.8 |
| MTUA | Pref. for task switching | 2.60 | 133 | 0.005 | 4.5** | 4.0 |

*Sig. diff. at the $p < 0.05$ level. **Sig. diff. at the $p < 0.01$ level.

pursuits, such as going out, developing a relationship with the user, and taking on emotional labour. This could represent higher-level differences in expectations between Eastern and Western cultures for media richness. As Huang and Zhang [5] found for Taiwan and UK participants when comparing transactional and nontransactional interactions, these results point to differences by culture based on expectations for dialogue, e.g., who asks a question, type of activity, and social context. Still, we might expect "predomestication" of VAs [82] to occur for both Japan and US respondents, given how machines are envisioned in Japanese and American popular culture. Given the kinds of entertainment suggested, we expected higher rates of certain activities among Japanese respondents, e.g., singing, which is prevalent in Japan (karaoke). These patterns and gaps should be explored in future work.

We summarize the forms of dialogue and activities in imagined VA interactions as follows:

(i) *Typical*: cooking, games, purchases, music, instructions, weather, news, reminders, appointments, time and timers, search queries, smart home management (e.g., lights, heat), directions, phone calls, and singing

(ii) *Radical*: coaching and advice, deep conversations, social presence for loneliness, access to another perspective, practice and feedback, jokes and comedy, offload childcare, advertisements, medical charting, use when other sensory modalities are inaccessible or when distracted (e.g., while driving), go into nature, talk therapy, mindfulness meditation, and "outside-the-home" activities

We summarize the overarching patterns in imagined VA interactions by country as follows:

(i) *American orientations*: transactional, exercise, and caretaking

(ii) *Japanese orientations*: conversation, serious pursuits, going out, embodied, emotional labour, and relational

6.2. Cultural Sensitivity and the Value of Cross-Cultural Comparisons

6.2.1. Cultural Sensitivity over WEIRD Sampling. Our cross-cultural approach revealed similarities as well as clear differences between Japan and the US. Similarities included chit-chat, recommendations, and opinions from the VA alongside more ideological engagement, humour, and even contradictions from the VA. Japanese and US respondents also presented the VA as technically omniscient and ever-connected. US respondents, however, tended to craft dialogues on personal topics, building a relationship with the VA and using it as a sounding board. They expected the VA to know about them and their environment, as well as take the lead and act on implicit next cues. Conversely, Japanese respondents imagined VAs as less personal but still knowledgeable: capable of thinking ahead, using omoiyori to take action without being asked, and being able to "read the room," or kūki wo yomu. They also presented the VAs as socially aware: knowledgeable about people other than the user. These findings highlight cultural differences in attitudes towards VAs. As an "EIRD" nation, Japan shows how culture influences technology *attitudes* despite education, access, and societal integration of technology. The implication is that a reliance on WEIRD sampling and the findings of work based on WEIRD sampling alone is risky. We may be able to crowdsource translations, but how people are oriented towards VAs and technology in their culture may lead to unexpected effects and raise new challenges for designers.

Table 7: Statistically significant correlations for dialogue subthemes and technology attitudes (MTUA).

| Country | Dialogue subtheme | $r$ | df | $rp$ | $\tau_b$ | $\tau_b p$ | US mean | JP mean | Sig. |
|---|---|---|---|---|---|---|---|---|---|
| Both | Interpersonal connections | 0.31 | 99 | 0.002 | 0.28 | 0.001 | 5.1 | 4.9 | ** |
| Both | Recommending | 0.24 | 99 | 0.02 | 0.24 | 0.004 | 5.2 | 5.0 | ** |
| Both | Giving an opinion | 0.23 | 99 | 0.02 | 0.23 | 0.006 | 5.2 | 4.8 | ** |
| Both | Thinking ahead | 0.18 | 99 | 0.07 | 0.19 | 0.03 | 4.6 | 5.4 | * |
| Both | Contradicting | 0.16 | 99 | 0.10 | 0.18 | 0.03 | 5.2 | 5.3 | * |
| Both | Giving the VA the lead | 0.18 | 99 | 0.07 | 0.18 | 0.03 | 5.1 | 4.7 | * |
| Both | Implicit next | 0.25 | 99 | 0.01 | 0.23 | 0.006 | 5.1 | 4.8 | ** |
| JP | Interpersonal connections | 0.28 | 42 | 0.06 | 0.27 | 0.04 | | 4.9 | * |
| JP | Thinking ahead | 0.46 | 42 | 0.002 | 0.38 | 0.003 | | 5.5 | ** |
| JP | About the user | 0.32 | 42 | 0.03 | 0.27 | 0.03 | | 5.1 | * |

∗Sig. diff. at the $p < 0.05$ level. ∗∗Sig. diff. at the $p < 0.01$ level. Based on the larger $p$ value.

Table 8: Statistically significant correlations for activity themes and technology attitudes (MTUA).

| Country | Activity theme | $r$ | df | $rp$ | $\tau_b$ | $\tau_b p$ | US mean | JP mean | Sig. |
|---|---|---|---|---|---|---|---|---|---|
| Both | Serious pursuits | 0.23 | 71 | 0.047 | 0.18 | 0.065 | 5.5 | 4.9 | * |
| Both | Caretaker | 0.28 | 71 | 0.02 | 0.23 | 0.02 | 5.4 | 5.0 | * |
| US | Serious pursuits | 0.34 | 32 | 0.048 | 0.30 | 0.047 | 5.5 | | * |

∗Sig. diff. at the $p < 0.05$ level. Based on the larger $p$ value.

Stronger technology attitudes in the US participants also played a key role. Unlike in Völkel et al. [15], this was linked to more turns within dialogues, presenting the VA as the questioner, and a desire for advanced social interaction, especially serious pursuits and the VA as a personal caretaker. US respondents also used wake words, the special commands to "wake up" the VA at the start of an interaction, which no Japanese respondent used. Still, technically savvy Japanese respondents used more plain language or tameguchi with the VA. They also wrote dialogues and offered activities similar to those in the US, especially interpersonal connections and knowing the user. These results reflect the relative prevalence of VAs in the US market compared to Japan. Americans may be primed for certain interactions, such as wake words, as well as ready for new interactions with VAs. Japanese consumers may not yet have the needed interest or experience with VAs. Notably, US respondents used far more grounding behaviours than Japanese respondents. We can interpret this as a learned response to experiences with VA failure. As a wealth of previous research has shown [1, 3, 20, 21], VA failures are common and disruptive. This finding is a sign that US respondents already have mental models of VAs being "prone to failure." We can design responses to grounding behaviour. But we can also explore novel solutions in the Japanese market, if we can avoid the failures that lead to grounding behaviour in the first place—a technical challenge that co-design work such as this can inform.

We summarize the key similarities and differences between US and Japanese respondents as follows:

(i) *American orientations:* social engagement and longer exchanges based in mutual respect, with the VA as a caretaker, sounding board, and advisor that knows the user and their environment and is able to take the lead

(ii) *Japanese orientations:* relational engagement alongside short exchanges, perhaps one-way and transactional interactions, including ones that involve emotional labour and advanced social abilities (omoiyori and kūki wo yomu), with the use of various forms of polite and plain language (tameguchi) from either partner

*6.2.2. Aligning the Design of VAs with Japanese Language and Culture.* New ways of creating dialogues for VAs, notably NLP, are aimed at including a diversity of languages and cultures. Yet, we must be careful about relying on translation, which can be inexact and lossy. While US respondents used more social protocols and tended to interact with the VA in a mutually respectful way, Japanese respondents used a variety of politeness levels for the user and the VA. Context, including the relative social power and psychological closeness of the conversation partners [71]; word use, such as plain language (tameguchi); and verb conjugation need to be considered. Echoing Ouchi et al. [56], we found that most dialogues (92 or 90%) involved plain language (tameguchi). Notably, plain language and politeness do not translate between English and Japanese very well. Reverse translating one of our examples is telling "What's the weather like on the weekend?", which may be directly translated into Japanese as "kon-shū-matsu no tenki wa"—good, but—"dou deshou ka?"—these "polite" add-ons are unexpected. Moreover, one-fifth of Japanese respondents switched politeness levels, even mid-dialogue. Future work asking participants about their awareness and intent would be illuminating. Slang was also culturally driven, with US respondents relying on social media slang and Japanese respondents using katakana for special references or effect. This requires localization rather than transliteration. Finally, Japanese respondents avoided harassing and gendered language. Yet, translators and translation software make assumptions. Google Translate and DeepL assume "male as default," such as by translating the gender-neutral "san" into "Mr." This is why generating dialogues directly from people can complement translation work. In fact, it may be necessary to reflect real user needs and desires from within a language or cultural group.

Cultural differences in expressions of interpersonal closeness can be explained by norms in Japan. Excessive or overly intimate self-disclosures among strangers are not well received [83]. Japan is also characterized as a *morally relativistic* culture, where a person's response may change based on their perception of local or societal norms [84]. An example is tatemae-honne. Honne refers to one's true feelings, often hidden to spare feelings or conform to social expectations. Tatemae is the self presented to others when honne is hidden. This can explain the differences found for advice. For Japanese people, the desired advice reflects a morally relativistic perspective in which the person asking for advice considers the other party first, rather than disclosing how they feel right away. This points to a new factor in customizing the design of voice UX to specific end-users. While most work has been conducted in the West and focused on personality customization (e.g., [15, 43, 64]), these findings suggest that language and social norms within specific cultural contexts need to be incorporated into the design process of dialogues and agent behaviour. Future work can continue to move beyond WEIRD contexts [10] with further cross-cultural comparisons and work outside of Western nations to tease out the roles of language, social norms, and moral codes in voice UX.

We offer the following patterns for VA design in the Japanese context:

(i) *Linguistic expressions*: plain and polite language, including slang, truncated sentences, verb forms, and plain language (tameguchi), as well as pluralistic forms of gendered language, especially "san" and "kun," need to be trained into VAs. Automated translators could provide a range of options based on plain and polite form and word choice; this plurality should thus be included in NLP data sets

(ii) *Dialogic structures*: conversational, with fewer turns, and emotional labour, such as disclosures and an implicit desire for commiseration (zannen) and advice-seeking

(iii) *Social norms*: considering others by withholding judgments and negative opinions (tatemae-honne), thinking ahead instead of asking directly (omoiyari), and reading the room (kūki wo yomu); i.e., the VA must be trained on NLP data sets that include a range of subtle, unstated social cues

*6.3. The Merits and Limits of Crowdsourcing Co-Design Processes Cross-Culturally.* Co-designing dialogues and activities generated a wealth of material that may confirm and refute designer expectations. Still, findings should be carefully considered in light of how we framed the VA as ideal and against what we know about user mental models of VAs [85]. Comparing by country revealed patterns related to familiarity with modern VAs. This is a strength of our cross-cultural approach: we can begin to identify, albeit in broad strokes, the mental models of groups of people more or less familiar with the VAs of today. Japan may present an opportunity to set cultural expectations of VAs. For instance, Japanese respondents imagined forms of interaction that extended into the social environment. The gaps are provocative: what if our VAs became friends with our friends' VAs? Could we trust the VA to act on our behalf if it knows us well—and would people accept our VA as a proxy? In short, experts can use this user-generated material to inform the design of future VAs.

Yet, content from nonexperts also requires a critical eye. We found instances of VA harassment; antisocial attitudes towards ethnic minorities, i.e., Zainichi Koreans; and gendered language. We cannot avoid VA gendering and negative interactions with VAs, including sexist, racist, and otherwise harmful content [37, 46, 72]. For NLP data sets, Strengers et al. [37] offer three approaches: remove sensitive information, even though this prevents an opportunity for prosocial engagement; guide the user away from gendered engagements; and purposefully queer gendered engagements by going against stereotypes. Chin et al. [86] suggest an empathetic approach to avoid user anger and guilt. Perhaps a multipronged approach is ideal, which future work can explore cross-culturally.

We offer the following merits, limits, and opportunities for co-designing VAs with nonexperts, paying special attention to crowdsourcing approaches and elicitation methods:

  (i) *Welcome Cultural Sensitivity*. Co-design processes can be scaled up with crowdsourcing. This includes bringing in participants from other language and cultural contexts. We encountered no detriments in our cross-cultural approach. If the team has expertise in each language or culture, it should be feasible and rewarding

  (ii) *Bring in Embodiment*. Imagination has its limits. Also, the anonymity that online crowdsourcing initiatives allow may be useful or not. People are free to write what they wish, even if it is negative or taboo. But VAs do not have questionnaire bodies. Future work will need to prototype and test VAs with people directly. As Clark et al. [22] note, we can use a range of low and high fidelity approaches, including Wizard of Oz [87], where a human pretends to be the VA

  (iii) *Embrace Deviancy*. We should invite "deviant" dialogues and activities. People will use VAs in unexpected ways. They may not be forthcoming with experts in traditional face-to-face co-design studies due to observer effects and social acceptability biases. Looking at raw transcripts from real users may be enlightening but ethically murky. Feminist HCI and social justice approaches to NLP have already started on this work [7, 37, 44, 46, 72, 86]. Harassment and other negative forms of engagement are one type. Humour, including prank calls and jokes, and radical ideas for VA behaviour were also imagined. Future work can explore whether these are generalizable

  (iv) *Be Critical*. Co-design work, especially crowdsourced and compared across cultures, can reveal how the average person understands VAs. Limitations in imagination, our materials, unintentional priming, groupthink, and other well-mapped phenomena can lead to a narrow set of ideas. But this can still be informative. Experts can be critical, draw on their knowledge, and run follow-up (co-)design work with specific ideas and prototypes that invert expectations, centre marginalized perspectives, and offer new ideas. For example, an expert could take one of the radical activities in this work and present it to experts and/or nonexperts as a design fiction [31, 88, 89]

6.4. *Limitations*. This work generated imagined and ideal visions of VAs that should be tested in user studies. This is an ongoing limitation of elicitation work [1, 15]. We encouraged longer elicitations to address concerns about short elicitations [1] but were not successful. We recommend using a word limit in the questionnaire. We also could not force a conversational structure in the online platform we used, resulting in ambiguous input for "you" and "your partner." Our data sets were sufficient for research purposes, but future work should add on for training VAs. Finally, the Japanese sample was older than the US sample, which could have influenced the results. Balanced samples and intergenerational work can clarify whether this is an issue.

## 7. Conclusions

Voice interaction with computer agents is increasingly found across languages and cultures. We have provided a co-designed or "coimagined" cross-cultural perspective to guide the design of future interactions with VAs. In this crowdsourced work, we have shown, by contrasting the English and Japanese contexts, how designing VAs must be done with cultural sensitivity. We add our voice to the call for moving beyond current forms of VA interaction and a reliance on WEIRD, or at least "W," populations in VA design and research.

## Data Availability

The data set is available via OSF at https://osf.io/bm8r2/.

## Conflicts of Interest

The authors declare no competing interests.

## Acknowledgments

We thank Mutsumi Kashiwabara for assisting with the translation of the research instruments, Tatsuya Itagaki for a theme recommendation, and Peter Pennefather for reviewing the draft of this manuscript. This work was funded by a Japan Society for the Promotion of Science (JSPS) Grant-in-Aid for Early-Career Scientists (KAKENHI Wakate) (21K18005). Open Access funding is enabled and organized by JUSTICE Group 1 2023.

## Supplementary Materials

Our supplementary material contains the English and Japanese questionnaires. *(Supplementary Materials)*

[34] A. A. Lewinski, R. A. Anderson, A. A. Vorderstrasse, and C. M. Johnson, "Developing methods that facilitate coding and analysis of synchronous conversations via virtual environments," *International Journal of Qualitative Methods*, vol. 18, 2019.

[35] C. N. Harrington, R. Garg, A. Woodward, and D. Williams, ""It's kind of like code-switching": black older adults' experiences with a voice assistant for health information seeking," in *CHI '22: Proceedings of the 2022 CHI Conference on Human Factors in Computing Systems*, pp. 1–15, New York, NY, USA, April 2022.

[36] R. Li, S. E. Kahou, H. Schulz, V. Michalski, L. Charlin, and C. Pal, "Towards deep conversational recommendations," in *Advances in Neural Information Processing Systems*, Curran Associates, Inc., 2018, August 2022, https://papers.nips.cc/paper/2018/hash/800de15c79c8d840f4e78d3af937d4d4-Abstract.html.

[37] Y. Strengers, L. Qu, Q. Xu, and J. Knibbe, "Adhering, steering, and queering: treatment of gender in natural language generation," in *CHI '20: Proceedings of the 2020 CHI Conference on Human Factors in Computing Systems*, pp. 1–14, Honolulu, HI, USA, April 2020.

[38] E. Bastianelli, A. Vanzo, P. Swietojanski, and V. Rieser, "SLURP: a spoken language understanding resource package," in *Proceedings of the 2020 Conference on Empirical Methods in Natural Language Processing (EMNLP)*, pp. 7252–7262, 2020.

[39] J. Ledlie, B. Odero, E. Minkov, I. Kiss, and J. Polifroni, "Crowd translator," *ACM SIGOPS Operating Systems Review*, vol. 43, no. 4, pp. 84–89, 2010.

[40] X. Wang, J. Wu, J. Chen, L. Li, Y.-F. Wang, and W. Y. Wang, "VaTeX: a large-scale, high-quality multilingual dataset for video-and-language research," pp. 4581–4591, 2019, August 2022, https://openaccess.thecvf.com/content_ICCV_2019/html/Wang_VaTeX_A_Large-Scale_High-Quality_Multilingual_Dataset_for_Video-and-Language_Research_ICCV_2019_paper.html.

[41] D. Bilal and J. K. Barfield, "Hey there! What do you look like? User voice switching and interface mirroring in voice-enabled digital assistants (VDAs)," *Proceedings of the Association for Information Science and Technology*, vol. 58, no. 1, pp. 1–12, 2021.

[42] C. Whang and H. Im, ""I like your suggestion!" The role of humanlikeness and parasocial relationship on the website versus voice shopper's perception of recommendations," *Psychology & Marketing*, vol. 38, no. 4, pp. 581–595, 2021.

[43] L. Reicherts, N. Zargham, M. Bonfert, Y. Rogers, and R. Malaka, "May I interrupt? Diverging opinions on proactive smart speakers," in *CUI '21: Proceedings of the 3rd Conference on Conversational User Interfaces*, pp. 1–10, New York, NY, USA, July 2021.

[44] C. Chin and M. Robison, *How AI Bots and Voice Assistants Reinforce Gender Bias. Brookings*, 2020, December 2021, https://www.brookings.edu/research/how-ai-bots-and-voice-assistants-reinforce-gender-bias/.

[45] G. Hwang, J. Lee, C. Y. Oh, and J. Lee, "It sounds like a woman: exploring gender stereotypes in south korean voice assistants," in *CHI EA '19: Extended Abstracts of the 2019 CHI Conference on Human Factors in Computing Systems*, pp. 1–6, New York, NY, USA, May 2019.

[46] Y. Strengers and J. Kennedy, *The Smart Wife: Why Siri, Alexa, and Other Smart Home Devices Need a Feminist Reboot*, MIT Press, 2021.

[47] C. Nass, J. Steuer, and E. R. Tauber, "Computers are social actors," in *CHI '94: Proceedings of the SIGCHI Conference on Human Factors in Computing Systems*, pp. 72–78, Boston, MA, April 1994.

[48] J. A. Rode, "Reflexivity in digital anthropology," in *CHI '11: Proceedings of the SIGCHI Conference on Human Factors in Computing Systems*, pp. 123–132, Vancouver, BC, May 2011.

[49] C. Bartneck and J. Hu, "Exploring the abuse of robots," *Interaction Studies*, vol. 9, no. 3, pp. 415–433, 2008.

[50] E. Byrne, *Swearing Is Good for You: The Amazing Science of Bad Language*, Profile Books, 2017.

[51] A. Gomes, D. Antonialli, and T. Oliva, *Drag queens and Artificial Intelligence: should computers decide what is 'toxic' on the Internet?*, InternetLab, 2019, August 2022, https://internetlab.org.br/en/news/drag-queens-and-artificial-intelligence-should-computers-decide-what-is-toxic-on-the-internet/.

[52] I. F. Ogbonnaya-Ogburu, A. D. R. Smith, A. To, and K. Toyama, "Critical race theory for HCI," in *CHI '20: Proceedings of the 2020 CHI Conference on Human Factors in Computing Systems*, pp. 1–16, Honolulu, HI, USA, April 2020.

[53] E. ( D.) Tunstall, "Decolonizing design innovation: design anthropology, critical anthropology, and indigenous knowledge," in *Design Anthropology: Theory and Practice*, W. Gunn, T. Otto, and R. C. Smith, Eds., pp. 232–250, Bloomsbury, London, UK, 2013.

[54] S. Andringa and A. Godfroid, "Sampling bias and the problem of generalizability in applied linguistics," *Annual Review of Applied Linguistics*, vol. 40, pp. 134–142, 2020.

[55] K. Seaborn and Y. Kim, ""I'm" lost in translation: pronoun missteps in crowdsourced data sets," in *CHI EA '23: Extended Abstracts of the 2023 CHI Conference on Human Factors in Computing Systems*, pp. 1–6, New York, NY, USA, April 2023.

[56] S. Ouchi, K. Mizumaru, D. Sakamoto, and T. Ono, "Should Speech Dialogue System Use Honorific Expression?," in *HAI '19: Proceedings of the 7th International Conference on Human-Agent Interaction*, pp. 232-233, New York, NY, USA, September 2019.

[57] N. L. Hoft, "Developing a cultural model," in *International Users Interface*, E. M. Galdo and J. Nielsen, Eds., John Wiley & Sons, 1996.

[58] R. Heimgärtner, *Intercultural User Interface Design*, Springer, 2019.

[59] K. Seaborn, S. Chandra, and T. Fabre, "Transcending the "male code": implicit masculine biases in NLP contexts," in *CHI '23: Proceedings of the 2023 CHI Conference on Human Factors in Computing Systems*, pp. 1–19, New York, NY, USA, April 2023.

[60] S. Villarreal-Narvaez, J. Vanderdonckt, R.-D. Vatavu, and J. O. Wobbrock, "A systematic review of gesture elicitation studies: what can we learn from 216 studies?," in *DIS '20: Proceedings of the 2020 ACM Designing Interactive Systems Conference*, pp. 855–872, New York, NY, USA, July 2020.

[61] V. Clarke, N. Hayfield, N. Moller, and I. Tischner, "Once upon a time…: story completion methods," in *Collecting Qualitative Data: A Practical Guide to Textual, Media and Virtual Techniques*, V. Braun, V. Clarke, and D. Gray, Eds., pp. 45–70, Cambridge University Press, Cambridge, 2017, November 2022, http://www.cambridge.org/gb/academic/subjects/psychology/psychology-research-methods-and-statistics/